\documentclass[%
 reprint,
 amsmath,amssymb,
 aps,
]{revtex4-2}

\usepackage{graphicx}% Include figure files
\usepackage{dcolumn}% Align table columns on decimal point
\usepackage{subcaption}
\usepackage[dvipsnames]{xcolor}
\usepackage{bm}% bold math
\begin{document}

%\preprint{APS/123-QED}

\title{Curved momentum space and finite Landau spectrum in $\kappa$-Minkowski spacetime}% Force line breaks with \\

\author{Adrián Huamán Vargas}
 \email{adrian.huaman@ufabc.edu.br}%Lines break automatically or can be forced with \\
\author{Vladislav Kupriyanov}%
 \email{vladislav.kupriyanov@gmail.com}
\affiliation{%
CMCC - Universidade Federal do ABC - Brazil
}%

\date{\today}% It is always \today, today,

\begin{abstract}
We derive the $\kappa$-Poincar\'e Casimir from the de Sitter geometry of momentum space and employ it as the dynamical constraint governing charged particles within the framework of Poisson gauge theory. The resulting formalism is applied to scalar and spin-$1/2$ particles in a constant magnetic field. Within the semiclassical Poisson-gauge framework, exact energy spectra are obtained without expanding in the deformation parameter $1/\kappa$. On the positive-energy branch in bicrossproduct momentum coordinates, identifying these coordinates with the physical momenta entering the gauge coupling imposes an upper bound on their spatial magnitude. This bound leads to a finite number of admissible Landau-level indices and hence to a Highest Landau Level (HLL). In the fermionic case, the truncation is spin dependent for the factorization adopted here, resulting in a polarized HLL. Possible implications of this ultraviolet truncation and the limitations of the framework are briefly discussed.
\end{abstract}
%'anomaly-related phenomena' perhaps shoulb be removed
\maketitle

%\tableofcontents

\section{\label{sec:level1}Introduction}
One of the longstanding motivations for studying noncommutative geometry is the expectation that the classical notion of spacetime may cease to be valid at sufficiently short distances. In several approaches to quantum gravity, the structure of spacetime is expected to acquire a noncommutative character, leading to modifications of both the kinematics and the dynamics of relativistic particles. Among the various models that have been proposed, $\kappa$-Minkowski spacetime occupies a distinguished role due to its close connection with deformed relativistic symmetries, quantum groups, and Doubly Special Relativity \cite{sym13060946, Majid_1994}. In this framework, spacetime coordinates satisfy Lie-algebra-type commutation relations characterized by a deformation scale $\kappa$, usually associated with a fundamental ultraviolet scale.\\\\
A remarkable feature of $\kappa$-Minkowski spacetime is that the deformation can be interpreted geometrically in terms of a curved momentum space \cite{KOWALSKI_GLIKMAN_2013}. The composition law of momenta becomes non-Abelian and the corresponding momentum manifold is identified with an $AN(3)$ group manifold, which can be embedded as a region of four-dimensional de Sitter space \cite{Arzano_2010,Arzano_2011,Nandi_2023}. Within this picture, the $\kappa$-Poincaré Casimir acquires a natural geometric interpretation and replaces the usual relativistic mass-shell condition. As a consequence, particle dynamics are modified in a way that directly reflects the geometry of momentum space.\\\\
Several aspects of particle dynamics in $\kappa$-Minkowski spacetime have been investigated previously, including deformed dispersion relations \cite{pachol2011,Aschieri_2017}. Nevertheless, the relation between the geometric origin of the $\kappa$-Casimir, gauge-invariant couplings, and the resulting quantum spectra remains incompletely understood. In particular, it is natural to ask how the choice of physical momentum coordinates manifests itself in the dynamics of charged particles and whether genuinely new spectral phenomena emerge from the resulting deformed dispersion relation.\\\\
In this work, we revisit particle dynamics in $\kappa$-Minkowski spacetime from the perspective of the $\kappa$-Poincaré Casimir and its geometric interpretation in de Sitter momentum space. Using the framework of Poisson gauge theory \cite{Kupriyanov_2021,KupriyanovSzabo_2021,Kurkov:2021kxa,kupriyanov2024,Sharapov_2024}, and in particular its application to $\kappa$-Minkowski spacetime \cite{Vitale_2021,Vitale_2023,Kurkov_2025,Abla_2025,kurkov2026}, we construct gauge-invariant classical and quantum dynamics for scalar and spin-$1/2$ particles. The resulting equations are applied to the Landau problem, for which energy spectra can be obtained exactly in $1/\kappa$ within the Poisson-gauge approximation. Previous analyses of Landau levels in $\kappa$-Minkowski have been performed in a different framework \cite{LLNCKappa}; here we derive the spectrum starting directly from the $\kappa$-Casimir. A central result is the emergence of a finite set of admissible Landau-level indices in the positive-energy bicrossproduct branch when these momentum variables are used in the electromagnetic coupling. We further discuss the spin dependence of this truncation and its possible implications for the structure of $\kappa$-deformed quantum theories.
\section{$\kappa$-Poincare}
The $\kappa$-Minkowski spacetime is a Lie-algebra-type noncommutative spacetime whose coordinate operators $\hat{x}^\mu$ satisfy the commutation relations
\begin{align}
[\hat{x}^0,\hat{x}^i]=\frac{i}{\kappa}\hat{x}^i,
\end{align}
with all remaining commutators vanishing. As a consequence of this deformation of the ordinary commutative spacetime structure, the standard Poincaré symmetry of special relativity is no longer preserved. Instead, it is replaced by a deformed symmetry structure, known as the $\kappa$-Poincaré algebra \cite{LUKIERSKI1991331}.\\\\
\subsection{Deformed momentum space}
In this subsection we followed the logic used in \cite{Kowalski_Glikman_2017} and present it in an illustrative narrative. To make the deformation explicit, it is instructive to examine the composition law for momenta. Consider the time-to-the-right ordered plane waves
\begin{align}
g_k(\hat{x}) := e^{ik_i\hat{x}^i}e^{ik_0\hat{x}^0}.
\end{align}
The product of two such plane waves, $g_k(\hat{x})g_q(\hat{x})$, differs from the commutative case. In ordinary spacetime, the commutativity of the coordinate operators implies a linear addition law for momenta. In the present setting, however, the coordinate operators satisfy nontrivial commutation relations, and the composition of plane waves must be computed using the Baker--Campbell--Hausdorff formula. One then obtains the deformed momentum addition law
\begin{align}
(k\oplus q)_0 &= k_0 + q_0,&
(k\oplus q)_i &= k_i + e^{-k_0/\kappa}q_i.
\end{align}
This composition law endows the set of plane waves with a group structure,
\begin{align}
g_k\cdot g_q = g_{k \oplus q},
\end{align}
where the operation is associative, admits the identity element $k_\mu=0$, and possesses an inverse for every element, given by
\begin{align}
g(k_0,k_i)^{-1}=g(-k_0,-e^{k_0/\kappa}k_i).
\end{align}
The group obtained above is a particular example of the semidirect product $\mathbb{R}\ltimes\mathbb{R}^3$, commonly denoted by $AN(3)$. A well-known result in group theory states that $AN(3)$ can be realized as a subgroup of the five-dimensional Lorentz group $SO(4,1)$. This embedding allows the generators of $\kappa$-Minkowski spacetime to be represented in terms of the generators of $SO(4,1)$.\\\\
Let $J_{AB}$ denote the generators of $SO(4,1)$ in the fundamental representation,
\begin{align}
[J_{AB}]^C_{\ D}=\delta^C_A \eta_{BD}-\delta^C_B \eta_{AD},
\end{align}
which satisfy the algebra
\begin{align}
[J_{AB},J_{CD}]
=&\eta_{BC}J_{AD}-\eta_{AC}J_{BD} \\
&-\eta_{BD}J_{AC}+\eta_{AD}J_{BC}.
\end{align}
The generators of $\kappa$-Minkowski spacetime are then defined as
\begin{align}
X_i&:=\frac{i}{\kappa}(J_{4i}+J_{0i}), & X_0&:=\frac{i}{\kappa}J_{40}.
\end{align}
Since the Lie group $AN(3)$ is generated by four independent generators, the associated momentum manifold must be four-dimensional. This manifold can be identified with a region of four-dimensional de Sitter space $dS_4$, defined by the hyperboloid
\begin{align}
dS_4=\left\{P_A\in\mathbb{R}^{4,1} \middle| P_0^2-P_i^2-P_4^2=-\kappa^2\right\},
\end{align}
where $P_A$ denote the coordinates of the ambient space $\mathbb{R}^{4,1}$. Since every element $\Lambda\in SO(4,1)$ acts linearly on $\mathbb{R}^{4,1}$ and preserves its metric, it leaves the de Sitter hyperboloid invariant. Moreover, the action of $SO(4,1)$ on $dS_4$ is transitive, so that for any point $O\in dS_4$,
\begin{align}
SO(4,1)\cdot O=dS_4.
\end{align}
A convenient choice is
\begin{align}
O=(0,0,0,0,\kappa).
\end{align}
The stabilizer subgroup of this point is
\begin{align}
\mathrm{Stab}(O)=SO(3,1),
\end{align}
from which it follows that
\begin{align}
dS_4 \simeq SO(4,1)/SO(3,1).
\end{align}
Furthermore,
\begin{align}
\dim dS_4
= \dim SO(4,1)-\dim SO(3,1)
= 4.
\end{align}
Therefore, the four generators of the $\kappa$-Minkowski algebra provide a natural parametrization of this manifold. An explicit parametrization of the de Sitter momentum manifold can be obtained from the matrix representation of the plane waves. The exponentiated generators are given by
\begin{align}
e^{ik_0 X^0}&=\begin{pmatrix}
\cosh\left(\frac{k_0}{\kappa}\right) & 0 & -\sinh\left(\frac{k_0}{\kappa}\right) \\
0 & I_3 & 0 \\
-\sinh\left(\frac{k_0}{\kappa}\right) & 0 & \cosh\left(\frac{k_0}{\kappa}\right)
\end{pmatrix},\\
e^{ik_i X^i}&=\begin{pmatrix}
1+\frac{\vec{k}^{2}}{2\kappa^2} & \frac{k_i}{\kappa} & 1-\frac{\vec{k}^{,2}}{2\kappa^2}\\
\frac{k_i}{\kappa} & I_3 & -\frac{k_i}{\kappa}\\
1+\frac{\vec{k}^{2}}{2\kappa^2} & -\frac{k_i}{\kappa} & 1-\frac{\vec{k}^{2}}{2\kappa^2}
\end{pmatrix}.
\end{align}
Acting with the group element $g_k=e^{ik_iX^i}e^{ik_0X^0}$ on the reference point
\begin{align}
O=(0,0,0,0,\kappa),
\end{align}
yields the parametrization
\begin{align}
P_0&=-\kappa\sinh\left(\frac{k_0}{\kappa}\right)-\frac{\vec{k}^{2}}{2\kappa}e^{k_0/\kappa}, \\
P_i&=k_i e^{k_0/\kappa}, \\
P_4&=\kappa\cosh\left(\frac{k_0}{\kappa}\right)-\frac{\vec{k}^{2}}{2\kappa}e^{k_0/\kappa}.
\end{align}
The above construction reveals a remarkable feature of $\kappa$-Minkowski spacetime: while spacetime becomes noncommutative, momentum space acquires the structure of a curved manifold, namely a de Sitter space. The deformed composition law of momenta is encoded in the group structure of $AN(3)$, providing a geometric interpretation of the deformation parameter $\kappa$. This relation between noncommutative spacetime and curved momentum space underlies the emergence of the $\kappa$-Poincaré symmetry algebra, to which we now turn.
\subsection{Casimir}
The deformation of momentum space is accompanied by a deformation of the Poincare symmetry algebra \cite{Arzano_2011}. The resulting symmetry structure is described by the $\kappa$-Poincaré algebra, whose generators include rotations $M_i$ and boosts $N_i$. A convenient realization of these generators can be obtained from the phase-space relations
\begin{align}
[\hat{x}^0,\hat{p}_0]&=i, &
[\hat{x}^i,\hat{p}_j]&=i\delta^i_{\ j},\label{Rkappa1}\\ [\hat{x}^0,\hat{p}_i]&=-\frac{i}{\kappa}\hat{p}_i, & [\hat{p}_\mu,\hat{p}_\nu]&=0, \label{Rkappa2}
\end{align}
together with the $\kappa$-Minkowski commutation relations. It was found a poly-differential representation for this realization as
\begin{align}
    \hat{x}^0&=x^0+\frac{i}{\kappa}x^i\,\partial_ i, & \hat{x}^i&=x^i, & \hat{p}^\mu=i\partial^\mu.\label{rep}
\end{align}
In the context of non-commutative theories, quantization is predominantly performed in terms of deformation quantization, where interesting results are found for the $\kappa$-Minkowski star product \cite{Pacho__2015}. However, with this particular representation satisfying \eqref{Rkappa1} and \eqref{Rkappa2}, is clear that for any non time dependent potential, quantization can be achieved by regular canonical quantization. Later, in this realization, the boost generators take the form
\begin{align}
\hat{N}_i=\hat{x}^i\left(\frac{\kappa}{2}\left(1-e^{-2\hat{p}_0/\kappa}\right) + \frac{\vec{p}^{2}}{2\kappa}\right) +\hat{x}^0\hat{p}_i,
\end{align}
while the rotation generators retain their standard expression,
\begin{align}
\hat{M}_i=\varepsilon_{ijk}\hat{x}^j\hat{p}_k.
\end{align}
These generators satisfy the defining commutation relations of the $\kappa$-Poincaré algebra. We have thus obtained the full set of generators of the $\kappa$-Poincaré algebra, $\hat{p}_\mu,\ \hat{M}_i,\ \hat{N}_i$. Having specified the symmetry algebra, we may now construct an operator that remains invariant under all of its generators. This operator is the Casimir of the $\kappa$-Poincaré algebra, denoted by $\mathcal{C}_\kappa$, and is defined by
\begin{align}
[\hat{p}_\mu,\mathcal{C}_\kappa]= [\hat{N}_i,\mathcal{C}_\kappa]= [\hat{M}_i,\mathcal{C}_\kappa]= 0.
\end{align}
In the representation introduced above, which corresponds to the bicrossproduct basis, the Casimir takes the explicit form found in \cite{Kowalski_Glikman_2017}
\begin{align}
\mathcal{C}_\kappa(p) = 4\kappa^2
\sinh^2\left(\frac{\hat{p}_0}{2\kappa}\right) - e^{\hat{p}_0/\kappa}\hat{\vec{p}}^{2}.
\end{align}
In the commutative limit $\kappa\to\infty$, this expression reduces to the usual Poincaré Casimir,
\begin{align}
\lim_{\kappa\to\infty}\mathcal{C}_{\kappa}= \hat{p}_\mu \hat{p}^\mu=-\Box,
\end{align}
namely the d'Alembertian operator. The Casimir therefore plays a central role in the dynamics of the theory, as it determines the deformed mass-shell condition
\begin{align}
\mathcal{C}_\kappa = m^2.
\label{MassShellKappa}
\end{align}
It is important to clarify that the functional form of the Casimir is basis dependent. For instance, one may consider the classical-basis generators $P_\mu^{\rm cl}$, whose Lorentz-sector commutation relations are undeformed and whose Casimir is
 \begin{align}
     \mathcal{C}_\kappa^{\rm{cl}}=(P_0^{\rm cl})^2-(\vec{P}^{\rm cl})^2.
 \end{align}
 The two bases describe the same Hopf-algebra structure, and their invariants are related explicitly by
 \begin{align}
     \mathcal{C}_\kappa^{\rm{cl}}=\mathcal{C}_{\kappa}+\frac{\mathcal{C}_{\kappa}^2}{4\kappa^2}.
 \end{align}
 Consequently, imposing Eq.~\eqref{MassShellKappa} gives $\mathcal{C}_\kappa^{\rm cl}=m^2+m^4/(4\kappa^2)$. The corresponding free scalar equations therefore have the same eigenstates after this redefinition of the mass parameter. This free-theory equivalence does not by itself establish equivalence after coupling to an electromagnetic field: the nonlinear basis transformation and the complete gauge-coupling prescription must then be transformed consistently.\\\\
 Although the Casimir has already been obtained algebraically, it also admits a natural geometric interpretation in terms of the de Sitter momentum space introduced above. Recall that the coordinates $P_A$ define the embedding of the de Sitter hyperboloid in $\mathbb{R}^{4,1}$, while the parametrization $(P_0,P_i,P_4)$ is expressed in terms of the plane-wave coordinates ($k_\mu$). Since the action of $AN(3)$ leaves $P_0^2-\vec{P}^2$ invariant, then $P_4$ is also invariant. Therefore any linear map of $P_4$ is an invariant, also it is expected a quantity such that in the commutative limit reduces to the usual $p_\mu p^\mu$. This quantity turns out to be $2\kappa(P_4-\kappa)$. Substituting the parametrization of the hyperboloid into the quantity $2\kappa(P_4-\kappa)$, one finds
\begin{align}
2\kappa(P_4-\kappa) = 4\kappa^2\sinh^2\left(\frac{k_0}{2\kappa}\right) - e^{k_0/\kappa}\vec{k}^{2}.
\end{align}
This expression coincides precisely with the $\kappa$-Poincare Casimir. Therefore,
\begin{align}
\mathcal{C}_\kappa = 2\kappa(P_4-\kappa),
\end{align}
showing that the Casimir can be interpreted geometrically as a function of the embedding coordinates of de Sitter momentum space. Moreover, this identification establishes the equivalence between the $k_\mu$ parameters arising from the plane-wave construction and the real number representation $p_\mu$ of $\hat{p}_\mu$ introduced through the phase-space realization,
\begin{align}
k_\mu = p_\mu .
\end{align}
Consequently, the mass-shell condition
\begin{align}
\mathcal{C}_\kappa(p)=m^2
\end{align}
acquires a geometric interpretation as a constraint on the de Sitter momentum manifold and determines the corresponding deformed dispersion relation. 
\subsection{Solutions for positive and negative energies}
\begin{figure}[h!]
    \centering
    \begin{subfigure}[b]{0.7\linewidth}
        \includegraphics[width=\linewidth]{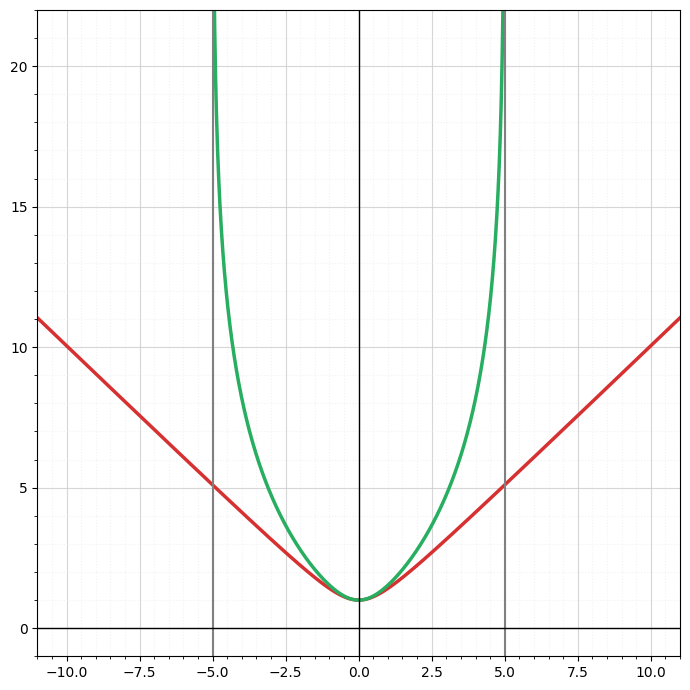}
        \caption{$E(|\vec{p}|)$}
        \label{Ep}
    \end{subfigure}
    \caption{Positive energy solutions $E(|\vec{p}|)$: the red line corresponds to the commutative limit, the green line to the $\kappa$-Minkowski energies and the grey lines to $|\vec{\pi}|=\kappa$.} 
    \label{Ekappa}
\end{figure}

Having established the deformed mass-shell condition that governs the dynamics of the theory, we now determine the corresponding energy spectrum, that is given by the quantity $p_0$ since $\hat{p}_0$ is the generator of time translation and $[\hat{p}_0,\hat{p}_0]=0$. Introducing the variable
\begin{align}
u:=e^{p_0/\kappa},
\end{align}
the Casimir relation can be rewritten as the quadratic equation\\
\begin{align}
(\kappa^2-\vec{p}^{2})u-(2\kappa^2+m^2)+\frac{\kappa^2}{u}=0.
\label{uEner}
\end{align}
Solving this equation for $u$, and consequently for the energy $E=p_0$, yields the two branches
\begin{widetext}
\begin{align}
E_{\pm}=\kappa \ln\left(\frac{1}{\kappa^2-\vec{p}^{2}} \left[\kappa^2+\frac{m^2}{2}\pm\sqrt{ \kappa^2(m^2+\vec{p}^{2}) +\frac{m^4}{4}}\right]\right).\label{EBranches}
\end{align}
\end{widetext}
As in the commutative theory, the deformed mass-shell condition gives rise to two distinct energy branches. At first sight, these solutions appear highly asymmetric, since in general $E_+\neq -E_-$. To understand the physical interpretation of the negative-energy branch, it is necessary to revisit the notion of antiparticles in the $\kappa$-Minkowski framework. In ordinary relativistic mechanics, a negative-energy solution with four-momentum $p_\mu^-$ is reinterpreted as a positive-energy antiparticle state through the transformation
\begin{align}
q_\mu=-p_\mu^-.
\end{align}
This construction relies on the fact that momentum addition is Abelian, so that the additive inverse of a momentum is simply its negative. In $\kappa$-Minkowski spacetime, however, the momentum composition law is deformed and the notion of inverse momentum must be modified accordingly. This leads to the introduction of the antipode map $S$, defined by
\begin{align}
p\oplus S(p)=0.
\end{align}
Using the deformed addition law, one finds
\begin{align}
S(p_0)&=-p_0,& S(p_i)&=-e^{p_0/\kappa}p_i.
\end{align}
The antipode is an anti-homomorphism of the symmetry structure and leaves the Casimir invariant,
\begin{align}
\mathcal{C}_\kappa(S(p))=\mathcal{C}_\kappa(p).
\end{align}
Consequently, negative-energy solutions can be reinterpreted as positive-energy antiparticle states by introducing the four-momentum
\begin{align}
q_\mu=S(p_\mu^-),
\end{align}
which generalizes the familiar relation $q_\mu=-p_\mu^-$ of the commutative theory. This way all solutions are taken to the positive branch and, for now on, it will be used exclusively this branch.\\\\
An immediate consequence of the positive-energy branch is that its domain is bounded from above,
\begin{align}
|\vec{p}|<\kappa.
\end{align}
On this positive-energy branch, the bicrossproduct spatial momentum approaches $\kappa$ only at infinite energy. This behavior is illustrated in Fig.~\ref{Ep}, which displays the energy as a function of $|\vec{p}|$ for $\kappa=5$ and $m=1$.\\\\
This upper bound is not a coordinate-independent consequence of the curvature of de Sitter momentum space: it depends on both the positive-energy branch and the bicrossproduct momentum coordinates. It acquires physical significance here because these variables are identified with the momenta entering the electromagnetic coupling. This identification is motivated by two features of the construction. First, the polydifferential representation \eqref{rep} gives $\hat{p}^\mu=i\partial^\mu$ directly, whereas other basis such as the classical-basis operators $\hat{P}_\mu^{\rm cl}(\partial)$ are nonlinear functions of derivatives. Second, as shown in Sec.~III, Poisson gauge theory naturally promotes the bicrossproduct variables to gauge-invariant momenta through the corresponding minimal-coupling prescription. The Landau-level bound derived below should therefore be understood as a result within this branch, basis, and coupling prescription, rather than as a basis-independent consequence of momentum-space curvature.
%whereas other basis such as the classical-basis...

\section{Charged particle dynamics}
\subsection{Poisson gauge theory}
To describe the coupling of a noncommutative theory to a vector field, we adopt the semiclassical framework of Poisson gauge theory \cite{Kupriyanov_2021}. In this approach, the full noncommutative gauge algebra,
\begin{align}
[\delta_f^{NC},\delta_g^{NC}]=\delta_{-iq[f,g]_\star}^{NC},
\end{align}
is replaced by its semiclassical counterpart, the Poisson gauge algebra,
\begin{align}
[\delta_f,\delta_g]=\delta_{q\{f,g\}},
\label{PGA}
\end{align}
where the Poisson bracket is obtained from the star-commutator through the correspondence principle of deformation quantization,
\begin{align}
\lim_{\hbar\to0}\frac{1}{i\hbar}[f,g]_\star =\{f,g\}.
\end{align}
In accordance with the realization used above to define $\hat{p}_\mu$, we introduce a symplectic embedding by extending the phase space with conjugate variables $p_\mu$. These variables play a central role in the construction of Poisson gauge theory. In the semiclassical limit, commutators are replaced by Poisson brackets, whose nonvanishing components are
\begin{align}
\{x^0,x^j\}&=\frac{1}{\kappa}x^j,&\{x^i,p_j\}&=\gamma^i_j =\delta^i_j, \label{PB1}\\\{x^0,p_0\}&=1,&  \{x^0,p_j\}&=\gamma^0_j= -\frac{1}{\kappa}p_j \label{PB2},
\end{align}
where $\gamma^\mu_\nu$ are the components of the symplectic embedding, that are required to reproduce
\begin{align}
\lim_{\kappa\to\infty}\gamma^\mu_\nu=\delta^\mu_\nu.
\end{align}
Upon quantization in the representation \eqref{rep}, the resulting momentum generators are identified with the bicrossproduct realization. These relations determine the Poisson structure and lead to the following representation of the Poisson gauge algebra \eqref{PGA}:
\begin{align}
\delta_f A_\mu=
\gamma^\nu_{\mu}(x,A)\partial_\nu f+q\{A_\mu,f\}.
\label{PGTrans}
\end{align}
 To construct a minimal coupling in Poisson gauge theory, it is necessary to introduce an invariant momentum $\pi_\mu$, satisfying $\delta_f\pi_\mu=0$, or applying the chain rule:
\begin{align}
    (\gamma^b_c(A)\partial_A^c +\gamma^b_c(p)\partial_p^c)\pi_a=0,\label{MomInv}
\end{align}
A further consistency requirement is that the commutative limit reproduces the standard gauge-invariant momentum, $\pi_\mu^{(0)}=p_\mu-qA_\mu$. A family of solutions to the $\kappa$-Minkowski consistency conditions has been obtained in \cite{Bruno}, and takes the form
\begin{align}
\pi_\mu = \left(\omega\rho_\mu^{\ \nu}(qA)+ (1-\omega)\rho_\mu^{\nu}(p)\right) (p_\nu - qA_\nu),
\label{KappaMomInv}
\end{align}
where the deformation matrix is defined as
\begin{align}
\rho(p) = \mathrm{diag}\left(1,e^{p_0/\kappa},e^{p_0/\kappa},e^{p_0/\kappa}\right),
\end{align}
and $\omega \in \mathbb{R}$ parametrizes the family of solutions. This construction provides a gauge-invariant momentum, which can then be used to implement the minimal-coupling prescription $p_\mu \to \pi_\mu$ and build a gauge-invariant action. It also motivates the identification of $p_\mu$ with the physical momenta in the present construction. These variables have a direct interpretation within the gauge construction and provide the natural momenta in which to formulate the Landau problem. Changing basis amounts to a nonlinear redefinition of the momentum generators. Equivalence of the resulting interacting descriptions would require transforming the full gauge-coupling prescription consistently and is not established by the free Casimir relation alone.
\subsection{Action with Grassmann variables}
A gauge-invariant action for a charged scalar particle has been constructed in \cite{Bruno}, and is given by
\begin{align}
S[x,p,\Lambda] =-\int d\tau \left[ \dot{p}_a \bar{\gamma}^a_{b}(p) x^b + \Lambda \phi\right], \label{BrunoAct}
\end{align}
where $\Lambda$ is a Lagrange multiplier enforcing the mass-shell constraint $\phi$ and $\bar{\gamma}$ is the inverse of the symplectic embedding. The latter plays the role of the Hamiltonian constraint of the system. In the commutative limit, it reduces to the standard relativistic form
\begin{align}
\phi_0 = \pi_\mu^{(0)} \pi^\mu_{(0)} - m^2.
\end{align}
To ensure gauge invariance, the mass-shell constraint must be expressed in terms of the gauge-invariant momentum, $\phi=\phi(\pi)$. We now extend the phase-space description to include spin degrees of freedom. Following Berezin and Marinov \cite{BEREZIN}, one introduces Grassmann variables $\xi^\mu$ satisfying the Poisson-Dirac algebra
\begin{align}
\{\xi^\mu,\xi^\nu\}&=-i\eta^{\mu\nu}, &\{\xi^\mu,\xi^5\}&=0, & \{\xi^5,\xi^5\}=i, \label{GVA}
\end{align}
which provide a classical pseudoclassical description of spin. In this framework, one can construct a particle action whose quantization yields the Dirac equation, together with its squared form corresponding to the Klein–Gordon equation with spin.
Merging the previous constructions with the Poisson gauge invariant action \eqref{BrunoAct} we get
\begin{widetext}
 \begin{align}
     S[x,p,\xi,\chi,\lambda]=-\int d\tau \left[\dot{p}_\mu\bar{\gamma}_\nu^\mu(p)x^\nu+\dfrac{i}{2}\xi^\mu\dot{\xi}_\mu-\dfrac{i}{2}\xi^5\dot{\xi}^5+\dfrac{i}{2}\chi T_1(\pi)+\lambda T_2(\pi)\right],
 \end{align}
\end{widetext}
 where $\lambda$ and $\chi$ are Lagrange multipliers enforcing the constraints of the system. The primary (Dirac-like) constraint is given by:
 \begin{align}
     T_1(\pi)=\xi^\mu \pi_\mu+\xi^5m,
 \end{align}
 while the secondary constraint is defined through $T_2=i\{T_1,T_1\}$, which yields:
 \begin{align}
     T_2(\pi)=\pi_\mu \pi^\mu-m^2+i\xi^\mu\xi^\nu\{\pi_\mu,\pi_\nu\}.
 \end{align}
 Quantization is implemented via the correspondence principle, which promotes phase-space Poisson brackets to commutators, $\{,\} \to -i[,]$,
while Grassmann variables are quantized by replacing Poisson brackets with anticommutators, $\{,\} \to -i[,]_+$. The corresponding operators $\hat{\xi}^\mu$ and $\hat{\xi}^5$, satisfying \eqref{GVA}, generate a Clifford algebra and can be represented in terms of Dirac matrices as
\begin{align}
\hat{\xi}^\mu=-\frac{i}{\sqrt{2}}\gamma^5\gamma^\mu,
\qquad
\hat{\xi}^5=\frac{i}{\sqrt{2}}\gamma^5.
\end{align}
In this quantization scheme, the constraints $T_1=0$ and $T_2=0$ become operator conditions on the matter field,
\begin{align}
\hat{T}_1\psi=0,\ \hat{T}_2\psi=0.
\end{align}
The first constraint yields the Dirac equation,
\begin{align}
i\gamma^5 \hat{T}_1 \psi= (\gamma^\mu \hat{\pi}_\mu - m)\psi = 0,
\end{align}
while the second leads to the squared Dirac equation (i.e. the Klein–Gordon equation with spin coupling),
\begin{align}
\hat{T}_2\psi=\left(\hat{p}_\mu \hat{p}^\mu - m^2 - i S^{\mu\nu}[\hat{p}_\mu,\hat{p}_\nu]\right)\psi= 0,
\end{align}
where $S^{\mu\nu}=\frac{i}{4}[\gamma^\mu,\gamma^\nu]$.
In the action constructed above, the constraints are expressed in terms of the invariant momentum $\pi_\mu$. This suggests that the gauge coupling can be implemented through the replacement $\hat{p}_\mu \rightarrow \hat{\pi}_\mu$, thereby ensuring gauge invariance at the quantum level. However, a consistent quantization of this prescription requires the explicit operator realization of $\hat{\pi}_\mu$ associated with the Poisson invariant momentum $\pi_\mu$, which in turn depends on the full structure of the underlying gauge algebra.
In the present work, we therefore restrict our analysis to the choice $\omega=1$ in \eqref{KappaMomInv}. In this case, and for configurations with $A_0=0$, the invariant momentum reduces to the standard minimal coupling form
\begin{align}
\pi_i = p_i - qA_i,
\end{align}
so that we expect the same structure to persist upon quantization.
 \subsection{$\kappa$-Minkowski action for spin 1/2 particle}
 We have shown how Grassmann variables can be introduced in order to construct a gauge-invariant action for a spin-$\tfrac{1}{2}$ particle. In this construction, the Dirac-like constraint $T_1$ is expressed in terms of both $p_\mu$ and the invariant momentum $\pi_\mu$, and the secondary constraint $T_2$ reproduces the standard mass-shell condition in the commutative limit. In $\kappa$-Minkowski spacetime, however, the mass-shell condition is deformed as a consequence of the curved geometry of momentum space. This makes the construction of the deformed constraint $T_1^\kappa$ nontrivial. To address this, we introduce new variables $\Pi_\mu$ such that the $\kappa$-Poincare Casimir can be written in the quadratic form
\begin{align}
C_\kappa = \Pi_\mu \Pi^\mu.
\end{align}
This factorization is not unique, and there is no unique choice of $\Pi_\mu$ satisfying $\lim_{\kappa\to\infty}\Pi_\mu=\pi_\mu$. Different choices correspond to inequivalent parameterizations of momentum space and may lead to different dynamical interpretations. For the purposes of this work, we adopt a particularly simple realization given by
\begin{align}
(\Pi_0)^2 &= 4\kappa^2\sinh^2\left(\frac{\pi_0}{2\kappa}\right),& (\vec{\Pi})^{2} &= e^{\pi_0/\kappa}\vec{\pi}^{2},
\end{align}
which reproduces the $\kappa$-Casimir in a direct and transparent way. The corresponding Dirac constraint in $\kappa$-Minkowski spacetime is therefore given by
\begin{align}
T_1^\kappa= \xi^\mu \Pi_\mu + \xi^5 m,
\end{align}
which leads to a deformed Dirac equation that is consistent with both $\kappa$-Poincaré symmetry and Poisson gauge invariance. Upon quantization, the constraint becomes the condition
\begin{align}
\left( 2\kappa \gamma^0 \sinh\left(\frac{\hat{\pi}_0}{2\kappa}\right) + \gamma^i:e^{\hat{\pi}_0/(2\kappa)} \hat{\pi}_i: -m\right)\psi=0,
\end{align}
where the colons denote an appropriate operator ordering prescription, which must be specified to ensure a consistent realization of the $\kappa$-deformed algebra. However, if $A_0=0$ and $\partial_t A_i=0$, there is no ordering problem. This construction of the Dirac equation differs from others such as \cite{verma2014,HARIKUMAR_2011}, which start from different mass-shell constraints and do not consider deformed gauge invariance. 
\section{Classical trajectories}
Having specified the deformed mass-shell constraint $\phi_\kappa$, we now investigate the associated classical dynamics in the presence of a constant magnetic field, as these trajectories provide insight into the corresponding quantum system. We consider the symmetric gauge vector potential
\begin{align}
\vec{A}&=\frac{B}{2}(-y,x,0),& A_0=0,
\end{align}
so that the invariant momentum reduces to the usual minimal coupling form $\pi_\mu = p_\mu - qA_\mu$. The evolution in proper time $\tau$ is generated by the constraint $\phi_\kappa$, such that the equations of motion for any phase-space function $f$ are given by
\begin{align}
    \frac{df}{d\tau}=\{f,\phi_k\},
\end{align}
with Poisson brackets given by \eqref{PB1} and \eqref{PB2}. This structure allows the derivation of the full equations of motion. For the specific choice of initial conditions $x_0=0$, $y_0=R$, $z_0=0$, $p_x^0=qBR/2$, $p_y^0=0$, $p_z^0=p_z$, the trajectories solutions reduce to
\begin{align}
    x(\tau)&=-R\sin(\theta(\tau)),\\
    y(\tau)&=R\cos(\theta(\tau)),\\
    z(\tau)&=\frac{\theta(\tau)}{qB}p_z,\\
    \theta(\tau)&=qBe^{p_0/\kappa}\tau,
\end{align}
where $R$ is the radius of oscillation. The structure of the trajectories coincides with the commutative limit $\kappa\to\infty$ except for the $e^{p_0/\kappa}$ in the angular parameter $\theta$.\\\\
The trajectories are bounded in the $xy$-plane, which supports the expectation of a discrete spectrum upon quantization. Also, it is known that the quantization of bounded symplectic manifolds is not trivial \cite{sharapov2026}. A second main consequence in the particular case of $\kappa$-Minkowski is the restriction imposed by the curvature of momentum space in this momentum realization:
\begin{align}
    \vec{\pi}^2=p_z^2+q^2R^2B^2<\kappa^2,
\end{align}
which implies that the parameters $R$ and $B$ are no longer independent. Since $B$ is an external field, this condition is interpreted as follows: for stronger magnetic fields, the maximal allowed radius of motion decreases, which in turn constrains the initial velocity of the particle. Equivalently, the total invariant momentum is distributed between motion along the $z$-direction and motion in the transverse plane.
\section{Landau problem and HLL}

\subsection{Spin 0} 
A particularly important example is the constant magnetic field problem, which provides the setting for the Landau level structure. In the context of $\kappa$-Minkowski spacetime, previous works such as \cite{LLNCKappa} have investigated the effects of a magnetic background on Landau levels; however, these approaches are not directly derived from the $\kappa$-Casimir structure. The formulation presented here therefore provides a complementary and Casimir-based derivation. Exact solutions can be obtained at all orders provided that $A_0=0$, since in this case the invariant momentum satisfies $\hat{\pi}_0=\hat{p}_0$, allowing stationary states of the form
\begin{align}
\psi(t,\vec{x}) = e^{-iEt}\psi(\vec{x}).
\end{align}
Under this assumption, the $\kappa$-deformed Klein–Gordon equation becomes
\begin{align}
\left(4\kappa^2\sinh^2\left(\frac{E}{2\kappa}\right)+ e^{E/\kappa}\vec{D}^{2}-m^2\right)\psi = 0,
\end{align}
where $\vec{D}$ denotes the covariant derivative in the magnetic background
\begin{align}
    \vec{D}=\nabla-iq\vec{A}.
\end{align}
Rearranging the terms, the equation can be written in a Schrödinger-like form,
\begin{align}
-\frac{\vec{D}^{2}}{2m}\psi=\frac{e^{-E/\kappa}}{2m} \left[4\kappa^2\sinh^2\left(\frac{E}{2\kappa}\right) -m^2\right]\psi.
\end{align}
This equation takes the form of a Schrödinger-like eigenvalue problem with a nontrivial energy-dependent coefficient. Since $\psi$ is an eigenstate of the covariant Laplacian,
\begin{align}
-\vec{D}^{2}\psi = (\Lambda_n)^2\psi, \label{LambdaN}
\end{align}
where $\Lambda_n$ are the corresponding eigenvalues. The equation reduces to an algebraic condition for the energy spectrum. The resulting energy coincides with the expression obtained previously in \eqref{EBranches}, with the invariant momentum replaced by the eigenvalues in Eq.~\eqref{LambdaN}:
\begin{widetext}
    \begin{align}
E=\kappa\ln\left( \frac{1}{\kappa^2-(\Lambda_n)^{2}}\left[ \kappa^2+\frac{m^2}{2} + \sqrt{\kappa^2(m^2+(\Lambda_n)^{2})+\frac{m^4}{4}} \right] \right).
\label{EnerK}
\end{align}
\end{widetext}
This agreement provides a consistency check between the $\kappa$-deformed Klein--Gordon formulation and the Hamiltonian analysis of the system. On the positive-energy bicrossproduct branch, physical states must satisfy
\begin{align}
|\vec{\pi}| < \kappa.
\end{align}
The denominator in Eq.~\eqref{EnerK} is therefore positive and the logarithm is well defined. This domain restriction follows from the chosen branch expressed in bicrossproduct momentum coordinates. Within this framework, once the stationary reduction is established, the method yields energy eigenvalues without an expansion in $1/\kappa$ for magnetic backgrounds of the type considered here. In particular, the $\kappa$-deformed dispersion relation remains algebraically solvable for a wide class of static vector potentials $\vec{A}$.

For the Landau problem, the operator $-\vec{D}^{2}$ reduces to a harmonic oscillator, and its eigenvalues are given by
\begin{align}
\Lambda_n^{2} = k_z^2 + (2n+1)|qB|.
\end{align}
The corresponding relativistic energy spectrum follows from the $\kappa$-deformed dispersion relation and takes the form
\begin{align}
\frac{E_n}{\kappa}=\ln\left(\frac{1}{\kappa^2-\Lambda_n^{2}}\left[\kappa^2+\frac{m^2}{2}+\sqrt{\kappa^2(\Lambda_n^{2}+m^2)+\frac{m^4}{4}}\right]\right).
\end{align}
Expanding to first order in $\kappa^{-1}$, the spacing between adjacent Landau levels is modified as
\begin{align}
\Delta E \approx \Delta E^{(0)} + \frac{|qB|}{\kappa}.
\end{align}
Finally, from the exact spectrum we recover the previously discussed kinematical bound $\vec{\pi}^{2}<\kappa^2$. In the present case this translates into a restriction on the quantum numbers,
\begin{align}
k_z^2 + (2n+1)|qB| < \kappa^2,
\end{align}
which limits the allowed range of Landau levels in the present branch and momentum realization. In contrast with the commutative case, where $n$ is unbounded for any fixed nonzero magnetic field, the bicrossproduct momentum bound restricts the admissible Landau-level indices. The resulting ultraviolet cutoff depends on the magnetic-field strength and on the longitudinal momentum. Because the inequality is strict, the highest accessible Landau level is
\begin{align}
n_{\mathrm{max}}=\left\lceil\frac{\kappa^2-k_z^2}{2|qB|}-\frac{1}{2}\right\rceil-1,
\end{align}
provided that the right-hand side is nonnegative. This expression, rather than the corresponding floor formula, also handles correctly the parameter values at which the upper bound is saturated. Increasing either $|k_z|$ or $|B|$ reduces the number of accessible Landau-level indices. It does not, by itself, remove the usual guiding-center degeneracy of each level.\\\\
Let $K=\kappa^2-k_z^2$. For $K>0$, only the lowest level $n=0$ survives in the interval
\begin{align}
\frac{K}{3}\leq |qB|<K.
\end{align}
For
\begin{align}
|qB|\geq K,
\end{align}
even the $n=0$ state is excluded and the stationary solutions on this branch cease to have real energy. This is not a fundamental upper bound on the external classical field; it is the condition for the existence of real-energy Landau states in the chosen branch and momentum realization. In terms of the magnetic length $l_B=1/\sqrt{|qB|}$, the existence of at least one scalar level requires $l_B>1/\sqrt{K}$, which reduces to $l_B>1/\kappa$ when $k_z=0$. The possible physical implications of the HLL are discussed at the end of this section.
\subsection{Spin 1/2}
For the fermionic case, we focus directly on the relativistic regime, since the non-relativistic limit follows analogously from the spin-0 analysis, where $\kappa$-deformation was shown to manifest primarily as a rescaling of the effective mass.
In the non-relativistic Pauli theory, the only additional contribution is the usual spin–magnetic coupling term $qB\sigma_z$, so that the energy spectrum retains the same structure as in the commutative case, with the replacement $m \to m_{\mathrm{eff}}$. In the relativistic case, however, additional care is required. As discussed in Section III.C, one does not obtain a unique Dirac-type constraint, but rather a family of constraints of the form
\begin{align}
T_1^\lambda = \xi^\mu \Pi_\mu + \xi^5 m,
\end{align}
whose secondary constraint yields the Casimir together with an additional mass-shell condition $ T_2^\lambda = i\{T_1^\lambda, T_1^\lambda\}.$
Explicitly, this reads
\begin{align}
T_2^\lambda = \Pi_\mu \Pi^\mu - m^2 + i\xi^\mu \xi^\nu \{\Pi_\mu, \Pi_\nu\}.
\end{align}
We adopt the following realization, which is the simplest choice consistent with both the Casimir structure and gauge invariance:
\begin{align}
\Pi_0 &= 2\kappa \sinh\left(\frac{\pi_0}{2\kappa}\right),&
\Pi_i &= e^{\pi_0/2\kappa}\pi_i.
\end{align}
This corresponds to a particular embedding of the $\kappa$-deformed phase space into an effective relativistic form. Different choices of realization are possible and may lead to nonequivalent dynamical extensions; a systematic classification of these alternatives lies beyond the scope of the present work. Within this choice, the $\kappa$-Dirac equation takes the form
\begin{align}
\left( 2\kappa \gamma^0 \sinh\left(\frac{\hat{\pi}_0}{2\kappa}\right) + \gamma^ie^{\hat{\pi}_0/2\kappa}\hat{\pi}_i- m \right)\psi = 0.
\end{align}
For the Landau problem, ordering ambiguities do not affect the final result in this background, and therefore no additional prescription is required. The secondary constraint takes the form
\begin{align}
\hat{T}_2 \psi= \left( \hat{\Pi}_\mu \hat{\Pi}^\mu
- m^2
- i S^{\mu\nu}[\hat{\Pi}_\mu,\hat{\Pi}_\nu]
  \right)\psi = 0.
  \end{align}
For a constant magnetic field oriented along the $z$-axis, with vector potential $\vec{A}=\frac{B}{2}(-y,x,0)$, the non-vanishing commutator becomes
\begin{align}
-i[\hat{\Pi}_x,\hat{\Pi}_y] = e^{\hat{p}_0/\kappa} qB.
\end{align}
Substituting this into the constraint yields the full $\kappa$-deformed Dirac–Landau equation,
\begin{align}
\left( 4\kappa^2 \sinh^2\left(\frac{\hat{p}_0}{2\kappa}\right) + e^{\hat{p}_0/\kappa}\vec{D}^{2}- m^2+ 2e^{\hat{p}_0/\kappa}qB S^z\right)\psi = 0,
  \end{align}
where the spin operator in the Dirac representation is given by
\begin{align}
  2S^z =\begin{pmatrix}
  \sigma_z & 0 \\
  0 & \sigma_z
  \end{pmatrix}.
\end{align}
To express the spectrum uniformly for either sign of $qB$, we define $s=\pm 1/2$ as the spin projection along $q\mathbf{B}$; equivalently, $s=\operatorname{sgn}(qB)s_z$, where $s_z$ is the eigenvalue of $S^z$. The spin term then modifies the eigenvalues from $\Lambda^2_n$ to $\Lambda^2_{n,s}$, where
\begin{align}
    \Lambda^2_{n,s}=k_z^2+(2n-2s+1)|qB|.
\end{align}
The spin-1/2 particle case is therefore closely related to the spin-0 case, with the replacement of the effective Landau parameter by its spin-dependent counterpart. For that, the complete energy spectrum can be obtained as before
\begin{widetext}
    \begin{align}
    E_{n,s}=\kappa\ln\left(\frac{1}{\kappa^2-(\Lambda_{n,s})^2}\left[\kappa^2+\frac{m^2}{2}+\sqrt{\kappa^2(m^2+(\Lambda_{n,s})^2)+\frac{m^4}{4}}\right]\right),
\end{align}
\end{widetext}
A further consequence of the momentum bound in this realization is a spin-dependent highest Landau level. The strict inequality $\Lambda_{n,s}^2<\kappa^2$ gives
\begin{align}
    n_{\mathrm{max}}(s)=\left\lceil\frac{\kappa^2-k_z^2}{2|qB|}+s-\frac{1}{2}\right\rceil-1,
\end{align}
provided that this integer is nonnegative. The spin-dependent term is specific to the factorization chosen for the Dirac constraint and should not be interpreted as a factorization-independent prediction. The scalar truncation, by contrast, follows directly from the Casimir within the chosen momentum branch and coupling prescription. The quantity entering the energy is the combined level
\begin{equation}
    N=n-s+\frac{1}{2}.
\end{equation}
Although a given value of $N$ may be represented by both spin projections through a balance of orbital and Zeeman contributions, the maximal orbital index differs by one between the two sectors. Consequently, whenever an admissible HLL exists, the state at the absolute maximum orbital index has $s=+1/2$ and is polarized along $q\mathbf{B}$. This detailed spin asymmetry is a property of the factorization adopted above.
\section{Discussion and conclusions}
One of the central lessons of $\kappa$-Minkowski spacetime is that the deformation does not merely modify the algebra of coordinates, but also changes the geometry of momentum space. In the present work, this geometric perspective was taken as the starting point. By identifying momentum space with an $AN(3)$ manifold embedded in de Sitter space, the $\kappa$-Casimir emerges naturally as a geometric invariant, providing the fundamental mass-shell condition that governs the dynamics of particles in the theory.\\\\
Using the framework of Poisson gauge theory, it was possible to construct gauge-invariant dynamics compatible with the $\kappa$-deformed symmetries. This allowed the formulation of both scalar and fermionic wave equations in the presence of electromagnetic fields. Although different realizations of the Casimir may lead to inequivalent fermionic theories, the particular realization adopted here provides a simple and consistent extension of the standard Dirac construction while preserving gauge invariance.\\\\
On the positive-energy branch expressed in bicrossproduct momentum coordinates, the mass-shell condition imposes the bound
\begin{align}
    |\vec{\pi}|<\kappa
\end{align}
when these coordinates are identified with the momenta entering the gauge coupling. This restriction is coordinate and branch dependent and is not a universal consequence of de Sitter momentum-space curvature. In the classical theory, it constrains the allowed trajectories and modifies the relation between energy and momentum. In the quantum theory, it leaves only finitely many admissible Landau-level indices. The corresponding HLL depends on the magnetic field and longitudinal momentum and, for the fermionic factorization adopted here, on the spin sector.\\\\
The truncation defines a projected physical subspace containing finitely many orbital indices, but it does not by itself imply a finite-dimensional transverse Hilbert space: on the infinite plane, each Landau level retains its guiding-center degeneracy. Moreover, the ordinary creation operator does not preserve the projected subspace, and imposing $a^\dagger|n_{\max}\rangle=0$ would require a projected or deformed ladder algebra rather than the standard Heisenberg--Weyl relation. Constructing such an algebra lies beyond the scope of the present work.\\\\
The meaning of ``exact'' in the present analysis requires qualification. The dependence on $1/\kappa$ is treated nonperturbatively: no expansion in $1/\kappa$ is made in deriving the displayed spectra. The gauge construction is nevertheless semiclassical because the full star-commutator algebra is replaced by its Poisson counterpart. The results are therefore exact to all orders in $1/\kappa$ within the Poisson-gauge approximation, but they do not constitute a complete deformation quantization of the interacting theory. In a fully noncommutative treatment, star-product and operator-ordering corrections may modify the gauge-covariant momentum algebra and hence the detailed Landau spectrum; whether the sharp truncation persists without modification remains an open question. For a general noncommutativity tensor $\Theta^{ab}(x)$, noncommutative gauge transformations may be constructed perturbatively, for example through the L$_\infty$ bootstrap and recurrence relations \cite{Kupriyanov:2023rrc}. Explicit all-order Poisson-gauge constructions are known for constant and linear (Lie--Poisson) structures and for certain nonlinear cases \cite{Kupriyanov_2021}.\\\\
Several aspects of the construction also deserve further investigation. The most important open problem concerns the non-uniqueness of the $\kappa$-deformed Dirac constraint. Different choices of quantities $\Pi_\mu$ reproduce the same $\kappa$-Casimir, and a systematic classification of the corresponding fermionic theories is still lacking. Likewise, a complete quantization of the gauge-invariant momentum $\pi_\mu$ would require a deeper understanding of the full non-commutative gauge algebra beyond the semiclassical Poisson description employed here.\\\\
In summary, within the positive-energy bicrossproduct branch and the Poisson-gauge coupling prescription adopted here, the $\kappa$-deformation modifies both the kinematics and the dynamics of relativistic particles. The resulting theory admits gauge-invariant scalar and fermionic dynamics, deformed dispersion relations, solutions exact in $1/\kappa$ within the semiclassical approximation, and a finite set of admissible Landau-level indices characterized by an HLL. These results illustrate how the choice of physical momentum variables in a curved momentum space can produce dynamical effects and make explicit which conclusions are realization dependent.

\section*{Acknowledgments}
We appreciate the support of the Funda\c{c}c\~ao de Amparo a Pesquisa do Estado de São Paulo
(FAPESP, Brazil), Grant numbers: 2024/04134 - 6 and 2024/00920-7. KVG also acknowledges financial support from the Concelho Nacional de Pesquisa (CNPq, Brazil), grant number: 305132/2024 - 5.


\begin{thebibliography}{99}

\bibitem{sym13060946}
M.~Arzano and J.~Kowalski-Glikman,
``An Introduction to $\kappa$-Deformed Symmetries, Phase Spaces and Field Theory,''
Symmetry \textbf{13} (2021) no.6, 946
doi:10.3390/sym13060946

\bibitem{Majid_1994}
S.~Majid and H.~Ruegg,
``Bicrossproduct structure of kappa Poincare group and noncommutative geometry,''
Phys. Lett. B \textbf{334} (1994), 348-354
doi:10.1016/0370-2693(94)90699-8

\bibitem{KOWALSKI_GLIKMAN_2013}
J.~Kowalski-Glikman,
``Living in Curved Momentum Space,''
Int. J. Mod. Phys. A \textbf{28} (2013), 1330014
doi:10.1142/S0217751X13300147

\bibitem{Arzano_2010}
M.~Arzano, J.~Kowalski-Glikman and A.~Walkus,
``Lorentz invariant field theory on kappa-Minkowski space,''
Class. Quant. Grav. \textbf{27} (2010), 025012
doi:10.1088/0264-9381/27/2/025012

\bibitem{Arzano_2011}
M.~Arzano and J.~Kowalski-Glikman,
``Kinematics of a relativistic particle with de Sitter momentum space,''
Class. Quant. Grav. \textbf{28} (2011), 105009
doi:10.1088/0264-9381/28/10/105009

\bibitem{Nandi_2023}
P.~Nandi, A.~Chakraborty, S.~K.~Pal, B.~Chakraborty and F.~G.~Scholtz,
``Symmetries of {\ensuremath{\kappa}}-Minkowski space-time: a possibility of exotic momentum space geometry?,''
JHEP \textbf{07} (2023), 142
doi:10.1007/JHEP07(2023)142

\bibitem{pachol2011}
A.~Pachol,
``$\kappa$-Minkowski spacetime: Mathematical formalism and applications in Planck scale physics,''
[arXiv:1112.5366 [math-ph]].

\bibitem{Aschieri_2017}
P.~Aschieri, A.~Borowiec and A.~Pacho{\l},
``Observables and dispersion relations in {\ensuremath{\kappa}}-Minkowski spacetime,''
JHEP \textbf{10} (2017), 152
doi:10.1007/JHEP10(2017)152

\bibitem{Kupriyanov_2021}
V.~G.~Kupriyanov,
``Poisson gauge theory,''
JHEP \textbf{09} (2021), 016
doi:10.1007/JHEP09(2021)016

\bibitem{KupriyanovSzabo_2021}
V.~G.~Kupriyanov and R.~J.~Szabo,
``Symplectic embeddings, homotopy algebras and almost Poisson gauge symmetry,''
J. Phys. A \textbf{55} (2022) no.3, 035201
doi:10.1088/1751-8121/ac411c

\bibitem{Kurkov:2021kxa}
M.~Kurkov and P.~Vitale,
``Four-dimensional noncommutative deformations of U(1) gauge theory and L$_{\infty}$ bootstrap.,''
JHEP \textbf{01} (2022), 032
doi:10.1007/JHEP01(2022)032

\bibitem{kupriyanov2024}
V.~G.~Kupriyanov, A.~A.~Sharapov and R.~J.~Szabo,
``Symplectic groupoids and Poisson electrodynamics,''
JHEP \textbf{03} (2024), 039
doi:10.1007/JHEP03(2024)039

\bibitem{Sharapov_2024}
A.~A.~Sharapov,
``Poisson electrodynamics with charged matter fields,''
J. Phys. A \textbf{57} (2024) no.31, 315401
doi:10.1088/1751-8121/ad62c7

\bibitem{Vitale_2021}
V.~G.~Kupriyanov, M.~Kurkov and P.~Vitale,
``$\kappa$-Minkowski-deformation of U(1) gauge theory,''
JHEP \textbf{01} (2021), 102
doi:10.1007/JHEP01(2021)102

\bibitem{Vitale_2023}
V.~G.~Kupriyanov, M.~A.~Kurkov and P.~Vitale,
``Lie-Poisson gauge theories and {\ensuremath{\kappa}}-Minkowski electrodynamics,''
JHEP \textbf{11} (2023), 200
doi:10.1007/JHEP11(2023)200

\bibitem{Kurkov_2025}
M.~A.~Kurkov,
``Light propagation in $\kappa $-Minkowski space-time: gauge ambiguities and invariance,''
Eur. Phys. J. C \textbf{85} (2025) no.10, 1231
doi:10.1140/epjc/s10052-025-14970-9

\bibitem{Abla_2025}
O.~Abla and M.~J.~Neves,
``Poisson electrodynamics on {\ensuremath{\kappa}}-Minkowski space-time,''
Phys. Lett. B \textbf{864} (2025), 139385
doi:10.1016/j.physletb.2025.139385

\bibitem{kurkov2026}
M.~Kurkov,
``Action principle for {\ensuremath{\kappa}}-Minkowski noncommutative U(1) gauge theory from Lie{\textendash}Poisson electrodynamics,''
J. Phys. A \textbf{59} (2026) no.14, 145402
doi:10.1088/1751-8121/ae5776

\bibitem{LLNCKappa}
M.~D.~{\'C}iri{\'c} and N.~Konjik,
``Landau levels from noncommutative U(1){\ensuremath{\star}} gauge theory in {\ensuremath{\kappa}}-Minkowski space-time,''
Int. J. Geom. Meth. Mod. Phys. \textbf{15} (2018) no.08, 1850141
doi:10.1142/S0219887818501414

\bibitem{LUKIERSKI1991331}
J.~Lukierski, H.~Ruegg, A.~Nowicki and V.~N.~Tolstoi,
``Q deformation of Poincare algebra,''
Phys. Lett. B \textbf{264} (1991), 331-338
doi:10.1016/0370-2693(91)90358-W

\bibitem{Kowalski_Glikman_2017}
J.~Kowalski-Glikman,
``A short introduction to $\kappa$-deformation,''
Int. J. Mod. Phys. A \textbf{32} (2017) no.35, 1730026
doi:10.1142/S0217751X17300265

\bibitem{Pacho__2015}
A.~Pacho{\l} and P.~Vitale,
``{\ensuremath{\kappa}}-Minkowski star product in any dimension from symplectic realization,''
J. Phys. A \textbf{48} (2015) no.44, 445202
doi:10.1088/1751-8113/48/44/445202

\bibitem{Bruno}
B.~S.~Basilio, V.~G.~Kupriyanov and M.~A.~Kurkov,
``Charged particle in Lie{\textendash}Poisson electrodynamics,''
Eur. Phys. J. C \textbf{85} (2025) no.2, 175
doi:10.1140/epjc/s10052-025-13897-5

\bibitem{BEREZIN}
F.~A.~Berezin and M.~S.~Marinov,
``Particle Spin Dynamics as the Grassmann Variant of Classical Mechanics,''
Annals Phys. \textbf{104} (1977), 336
doi:10.1016/0003-4916(77)90335-9

\bibitem{verma2014}
R.~Verma,
``Dirac Equation in $\kappa$-Minkowski space-time,''
[arXiv:1410.8680 [hep-th]].

\bibitem{HARIKUMAR_2011}
E.~Harikumar and M.~Sivakumar,
``$\kappa$-deformed Dirac Equation,''
Mod. Phys. Lett. A \textbf{26} (2011), 1103-1115
doi:10.1142/S021773231103550X

\bibitem{sharapov2026}
A.~A.~Sharapov,
``Quantization of bounded symplectic domains associated with compact Lie groups,''
J. Phys. A \textbf{59} (2026) no.2, 025203
doi:10.1088/1751-8121/ae3409

\bibitem{Kupriyanov:2023rrc}
V.~Kupriyanov, F.~Oliveira, A.~Sharapov and D.~Vassilevich,
``Non-commutative gauge symmetry from strong homotopy algebras,''
J. Phys. A \textbf{57} (2024) no.9, 095203


\end{thebibliography}
\end{document}